\documentclass[%
    reprint,
    superscriptaddress,
 amsmath,amssymb,
 aps,
]{revtex4-2}

\usepackage[normalem]{ulem}
\usepackage{siunitx}
\usepackage{graphicx}%
\usepackage{dcolumn}%
\usepackage{microtype}

\usepackage{bm}%
\usepackage{graphicx} 
\usepackage[dvipsnames]{xcolor}
\usepackage{todonotes}

\usepackage{hyperref}
\hypersetup{
    colorlinks=true,
    linkcolor=blue,
    citecolor=red,
    filecolor=magenta,      
    urlcolor=cyan,
}

\newcommand{\ud}{\mathrm{d}}

\newcommand{\nil}{null}

\let\vec\boldsymbol

\begin{document}

\title{Novel Signatures of Radiation Reaction in Electron--Laser Sidescattering}%

\newcommand{\HIJ}{Helmholtz Institute Jena, Fröbelstieg 3, 07743 Jena, Germany}
\newcommand{\GSI}{GSI Helmholtzzentrum für Schwerionenforschung GmbH, Planckstraße 1, 64291 Darmstadt, Germany}
\newcommand{\IOQ}{Institute of Optics and Quantum Electronics, Friedrich-Schiller-Universität, Max-Wien-Platz 1, 07743 Jena, Germany}
\newcommand{\CUOS}{G\'{e}rard Mourou Center for Ultrafast Optical Science, University of Michigan, 2200 Bonisteel Boulevard, Ann Arbor, Michigan 48109, USA}
\newcommand{\Berkeley}{Lawrence Berkeley National Laboratory, Berkeley, California 94720, USA}

\author{Philipp Sikorski}
\email{philipp.sikorski@uni-jena.de}
\affiliation{\IOQ}
\affiliation{\HIJ}
\affiliation{\GSI}

\author{Alec G. R. Thomas}
\affiliation{\CUOS}

\author{Stepan S. Bulanov}
\affiliation{\Berkeley}

\author{Matt Zepf}
\affiliation{\IOQ}
\affiliation{\HIJ}
\affiliation{\GSI}

\author{Daniel Seipt}
\email{d.seipt@hi-jena.gsi.de}
\affiliation{\GSI}
\affiliation{\HIJ}
\affiliation{\IOQ}

\date{\today}%

\begin{abstract}
In this article we investigate novel signatures of radiation reaction via the angular deflection of an electron beam colliding at 90 degrees with an intense laser pulse. Due to the radiation reaction effect, the electrons can be deflected towards the beam axis for plane wave backgrounds, which is not possible in the absence of radiation reaction effects. The magnitude and size of the deflection angle can be controlled by tailoring the laser pulse shapes. The effect is first derived analytically using the Landau-Lifshitz equation, which allows to determine the important scaling behavior with laser intensity and particle energy. We then move on to full scale 3D Monte Carlo simulations to verify the effect is observable with present day laser technology. We investigate the opportunities for an indirect observation of laser depletion in such side scattering scenarios.
\end{abstract}

\maketitle

    \section{Introduction}

    The phenomenon of (electromagnetic) radiation reaction (RR) describes the fact that accelerated charged particles emit radiation that carries away energy and momentum and, hence, the radiation emission must act back onto the motion of the particles. A correct treatment of particle dynamics thus must include the back-reaction of the radiation on the motion of the particles \cite{jackson}.
    Since the beginning of the 20th century, a number of different equations have been proposed to properly describe RR effects \cite{Lorentz1909,Abraham,Dirac:ProcRoySoc1938,Mo:PRD1971,Herrera:PRD1977,Ford:PLA1993,Landau,Ilderton:PLB2013} and novel models of RR are still being suggested \cite{Burton:ContempPhys2014,Burton:PLA2014,Noble:JMP2013,torgrimsson_resummation_2021}{, as well as models without self-interaction \cite{gratus_maxwelllorentz_2022}.} From a practical standpoint the equation of Landau and Lifshitz (LL) \cite{Landau} has gained some particular popularity due to its conceptual simplicity and ease of use in numerical calculations. The Landau-Lifshitz equation follows by a `reduction-of-order' from Lorentz-Abraham-Dirac equation \cite{ekman_exact_2021,ekman_reduction_2021,ekman_reduction_2022}.

    An often studied phenomenon of radiation reaction is the loss of energy of high-energy particles interacting with strong electromagnetic fields, such as ultra-strong laser fields \cite{Cole:PRX2018,Poder:PRX2018}. This is often discussed as beam cooling for electron-beam laser collisions \cite{Neitz:PRL2013,yoffe_longitudinal_2015,bulanov_energy_2024} and can even lead to population inversions in strong electromagnetic fields \cite{bilbao_radiation_2023}. The effect of electron-beam properties on laser-based radiation reaction experiments was studied in detail in Ref.~\cite{magnusson_effect_2023}.

    Radiation reaction effects can also occur in the quantum regime if the energy loss of an electron per hard photon emission is a significant fraction of their primary energy. Quantum effects become important if the quantum efficiency parameter $\chi = e\sqrt{p_\mu F^{\mu\nu} F_{\nu \lambda}p^\lambda}/m^3$ {approaches unity \cite{Ritus:JSLR1985,DiPiazza:RevModPhys2012,gonoskov_charged_2022,fedotov_advances_2023}}. Here $m$ is the electron mass, $e$ is the elementary charge, $p^\mu$ is the electron's four momentum and $F^{\mu\nu}$ is the electromagnetic field strength tensor. In the quantum regime, the stochastic nature of photon emission becomes a significant factor which causes a spreading (heating) of the particle's energy distributions \cite{Neitz:PRL2013,niel,Ridgers:JPP2017,zhang_quantum_2023,bulanov_energy_2024}.
    {Electron--laser collisions in this regime have been proposed as an opportunity to generate bright $\gamma$-ray flashes \cite{nakamura_high-power_2012,magnusson_laser-particle_2019,PhysRevLett.124.014801,gammaflash1}.}
    Analytical solutions for moments of the particle distribution have been recently found in Ref.~\cite{blackburn2023analytical,bulanov_energy_2024}. The influence of the electron spin on quantum radiation reaction effects has recently  been studied from kinetic equations in Ref.~\cite{seipt_kinetic_2023}. 

    In the quantum radiation reaction regime one expects also additional angular spreading \cite{green_transverse_2014,li_angle-resolved_2017,hu_quantum-stochasticity-induced_2020,beaming,bulanov_energy_2024} accompanying the particle heating. The control of the electron deflection in head-on collisions due to radiation reaction was proposed in Refs.~\cite{Heinzl:2013,Tamburini:PRE2014}. An efficient numerical code for calculating the final particle distributions in electron-laser collision has been made available \cite{amaro2023qscatter}.
    In a series of papers a novel matrix-differential equation for radiation reaction effects has been derived by explicit resummation of the underlying strong-field QED processes including polarization effects \cite{torgrimsson_resummation_2021,torgrimsson_resummation_2021b,torgrimsson_quantum_2023}.

    In this paper we identify a novel signature of radiation reaction, where an electron beam is deflected by radiation reaction in a 90 degree side-scattering scenario. The direction and magnitude of the deflection depend on the details of the interaction and can be in principle controlled by the laser pulse shape. The paper is organized as follows: In Sect.~\ref{sect:theory} we derived the radiative deflection in the side-scattering geometry from analytic solutions of the Landau-Lifshitz equation and find useful scaling laws for the magnitude of the effect. {In Sect.~\ref{sect:pulseshapes}~we} investigate several classes of laser pulse shapes that can provide suitable variability to control the direction and magnitude of the electron deflection. In Sect.~\ref{sect:sim} we present the results of QED Monte Carlo PIC simulations that show the robustness of the set-up with regards to realistic laser focusing and geometric overlap of the two beams. We summarize our work in Section \ref{sect:summary}.

    Throughout this work we employ rationalized Heaviside Lorenz units with $\hbar=c=\epsilon_0=1$. In these units the fine-structure constant is $\alpha=e^2/(4\pi)$, where $e>0$ is the elementary charge. Minkowski space scalar products are denoted in short-hand as $k.x\equiv k_\mu x^\mu$.

    \section{Theoretical Derivation of the Radiation Reaction Induced Electron Scattering Angle} \label{sect:theory}

    Let us begin our investigation by recalling the classical dynamics of an electron in an electromagnetic plane wave laser pulse in the absence of radiation reaction effects. For this case, the solution of the Lorentz force equation $\ud u^\mu/\ud\tau = -\frac{e}{m} F^{\mu\nu} u_\nu$ is well known, and reads
    \begin{align} \label{eq:umu}
        u^\mu = u^\mu_0 {+} a^\mu {-} k^\mu \frac{a.u_0}{k.u_0} {-} k^\mu \frac{a.a}{2k.u_0} \,,
    \end{align}
    where $u^\mu$ is the electron four-velocity, $\tau$ is the proper time, and $a^\mu(\phi)$ is the normalized vector potential of the plane wave, related to the field strength tensor via
    \begin{align}
        F^{\mu\nu} = \frac{m}{e} \left(k^\mu \frac{\ud a^\nu}{\ud\phi} - k^\nu \frac{\ud a^\mu}{\ud\phi}\right) \,,
    \end{align}
    with the laser wave-vector $k^\mu$ and phase $\phi=k.x$. To be specific, we take the plane wave propagating along the $z$-axis such that the light-front component $k^+$ is the only non-vanishing component of $k^\mu$.
    
    If we consider non-unipolar fields $a^\mu(-\infty) = a^\mu(\infty)$ without a memory effect \cite{dinu_infrared_2012,Ilderton2020,fedotov_advances_2023}, Eq.~\eqref{eq:umu} shows that the electron velocity after interaction with the plane wave is exactly the initial value before the interaction $u^\mu_0$. This is a manifestation of the Lawson-Woodward theorem stating that an electron cannot gain energy from interacting with a plane wave \cite{LW1,LW2,lawson_lasers_1979}, see also \cite{troha_reply_2002} and references therein. The light-front component $u^-(\phi) = u^-_0$ is a constant of motion \cite{harvey_symmetry}.

    The effect of the radiation reaction on the electron motion can be described using the Landau-Lifshitz equation \cite{Landau},
    \begin{multline} \label{eq:LL}
        \frac{\mathrm {d}u^\mu }{\mathrm {d}\tau }=  - \frac{e}{m} F^{\mu \nu } u_\nu 
        + {\tau_R \left[ -\frac{e}{m}(u^\alpha \partial_\alpha F^{\mu \nu }) u_\nu \right.}  \\
        { \left. 
        + \frac{e^2}{m^2}F^{\mu \nu } F_{\nu \alpha } u^\alpha 
        - \frac{e^2}{m^2} ( u_\alpha F^{\alpha \beta } F_{\beta\nu} u^\nu ) u^\mu 
        \right] } \,,
    \end{multline}
    {where $\tau_R= e^2/6\pi m$.}
    Exact analytical solutions are also known for Eq.~\eqref{eq:LL} in a plane electromagnetic wave \cite{Bulanov:PRE2011,piazza,blackburn,harvey_symmetry}.

    The electromagnetic field of a plane wave can be described by the normalized vector potential $a_\mu(\phi) = a_0 \sum_{i} \varepsilon_\mu^i f_i(\phi)$, with the polarization vectors $\varepsilon^{1,2}_\mu$ that satisfy $\varepsilon^i\cdot \varepsilon^j=-\delta_{ij}$ and $k\cdot\varepsilon^i=0$. The functions $f_j(\phi)$ describe the shape of the vector potential {wave form}. The condition of non-unipolarity requires that $f_j(-\infty) = f_j(+\infty)$.

    The exact solution of Eq.~\eqref{eq:LL} for the component $u^-$ can be given as \cite{Koga:PhysPlas2005}
    \begin{equation}
        u^-(\phi)= \frac{u^-_0}{1 + \frac{2}{3} R_{c} \mathcal I(\phi )},
        \label{eq:uminus}
    \end{equation} 
    where $u_0^-$ is the initial lightfront momentum, $R_c$ the classical radiation reaction parameter, which is defined as
    \begin{equation}
        R_{c} = %
        \frac{\alpha a_0^2 \: k.u_0 }{m} \,,
        \label{R}
    \end{equation}
    and
    \begin{align}
       \mathcal I(\phi )= \int _{-\infty }^\phi [f_1'(\psi )^2 + f_2'(\psi )^2]\,\ud\psi \,
       \label{Iphi}
    \end{align}
    is the (normalized) integrated laser intensity{---the radiant fluence---}seen by the electron, and the primes denote derivatives with respect to the laser phase $\phi$.

    Due to radiation reaction the value of $u^-$ is no longer conserved, but rather {monotonically} decreases as a function of $\phi$. In the absence of radiation reaction effects, $R_c\to0$, from Eq.~\eqref{eq:uminus} we recover light-front momentum conservation. Most detection schemes for radiation reaction are based on observing this (or a similar) radiative energy loss after the laser pulse has passed $\phi\to\infty$. In the following we propose as another observable for radiation reaction effects the angular deflection of the electrons.

    For the velocity components perpendicular to the wave propagation the analytical solution of Eq.~\eqref{eq:LL} can be given as \cite{piazza,blackburn,harvey_symmetry}
    \begin{multline}
        u^i(\phi) = \frac{1}{1 + \frac{2}{3} R_{c} \mathcal I(\phi )}\left [ u^{i}_{0} + a_0 f_i(\phi )+ \vphantom{\frac{1}{2}}\right.\\ 
        \left. + \frac{2 R_{c}}{3} a_0 \mathcal H_i(\phi )
        + \frac{2 R_{c}}{3 a_0} f_i'(\phi ) \right] \,
    \end{multline}
    where $u^{i}_{0}$ is the initial value of the perpendicular momentum component $i=\{1,2\}$, and $\mathcal H_i(\phi )$ is defined as
    \begin{align}
        \mathcal H_i(\phi ) = \int _{-\infty }^\phi \! f_i'(\psi )\,  \mathcal I(\psi ) \,\mathrm {d}\psi \,.
        \label{eq:Hphi}
    \end{align}
    From this one can get as a novel signature of radiation reaction an electron deflection if $\mathcal H_i(\phi \to \infty)$ is non-vanishing. {This happens if the laser pulse wave form possesses some strong asymmetry on the wave length scale.}
    We can integrate Eq.~\eqref{eq:Hphi} by parts to obtain
    \begin{align}
        \mathcal H_i(\phi) = f_i(\phi) \mathcal I(\phi) - \int_{-\infty}^\phi \! \ud\psi \: f_i(\psi) [f_1'(\psi)^2 + f_2'(\psi)^2] \,.
    \end{align}
    {In the following we will consider pulses with compact support.} At the end of the pulse, $\phi=\phi_f$, the first term vanishes. Since the term in the square brackets is strictly positive we need to maximize the asymmetry of the vector potential shape functions $f_i$ for large values of $\mathcal H_i$.

    To make the calculation of the electron deflection more concrete, let us consider a $90 \unit{\degree}$ scattering geometry in which the initial electron propagates along the $x$-axis. {Overall, $90 \unit{\degree}$ collisions have attracted much less attention in the literature than head-on collisions \cite{yoffe_electron_2017,vranic_multi-gev_2018}.
    For us, this geometry} seems favorable over the usually considered head-on collisions since the scattered electron directions will be nicely separated from the laser beam axis making the deflection angle detection much easier. We will also focus on linear laser polarization along the $x$-axis, i.e.~we will now set $f_2=0$.

    To find the electron properties after the interaction, we need to evaluate the velocity at the end of the pulse $u_f^\mu =u^\mu(\phi_f)$,
    \begin{align}
         u^{-}_{f} &=\frac{\gamma_0}{1 + \frac{2}{3} R_{c} \mathcal I(\phi_{f})} \,,  \\
         u^x_f &= \frac{1}{1 + \frac{2}{3} R_{c} \mathcal I(\phi_{f})} \left[ u^{1}_{0} + \frac{2 R_{c}}{3} a_0 \mathcal H_1 
         (\phi_{f}) \right] \,,\\
         u^y_f  & =0 \,,
    \end{align}
    where $u^\mu_0 = (\gamma_0, u_0^x,0,0)$ with initial Lorentz factor $\gamma_0$, $u_0^x=\sqrt{\gamma_0^2-1}$ and $u_0^-=\gamma_0$.

    The novel signature of radiation reaction is in a deflection angle upstream or downstream of the laser, $\tan \theta = {u^z}/{u^x}$. To predict the value of $\theta$ we need to calculate the component $u^z_f = (u^+_f  - u^-_f )/2$ along the laser beam axis, where $u^+$ is determined via the mass shell condition $u^+u^- - \vec u_\perp^2=1$, yielding
    \begin{align} \label{eq:uzf}
        u_f^z 
            &= \frac{\frac{2}{3}R_c( \mathcal I + \mathcal H_1 a_0  u^{x}_{0})+\frac{2}{9}R_c^2(\mathcal I^2 + \mathcal H_1^2a_0^2)}{\gamma_0(1+\frac{2}{3}R_c \mathcal I)} \,,
    \end{align}
    which is nonzero only due to radiation reaction effects. In the absence of the latter, $R_c\to 0$, we obtain $u^z_f=u^z_0=0$, consistent with the Lawson-Woodward theorem. Here it is understood that $\mathcal I$ and $\mathcal H_1$ are evaluated at $\phi_f$.
    With Eq.~\eqref{eq:uzf} we find for the deflection angle
    \begin{align} \label{eq:angle}
    \tan \theta = \frac{\frac{2}{3}R_c( \mathcal I + \mathcal H_1 a_0 u^x_0)+\frac{2}{9}R_c^2( \mathcal I^2+\mathcal H_1^2a_0^2)}{\gamma_0(u^x_{0}+\frac{2 R_{c}}{3} a_0 \mathcal H_1)} \,.
    \end{align}

    Let us now estimate the order of magnitude of $\theta$ as a function of the electron and laser parameters. For the $90\unit{\degree}$ scattering geometry we have $R_c = \alpha a_0^2 \omega_0 \gamma_0/m$, where $\omega_0$ is the laser frequency. It is now convenient to define as auxiliary constant $K=\alpha \omega_0/m$, which is the inverse of the classical critical vector potential $K^{-1} = a_\mathrm{crit}= m /\alpha \omega_0$. For a laser with $800 \, \unit{\nano\metre}$ wavelength we have $K\approx2.2\times 10^{-8}$. If we will take $a_0$ in the range $a_0\lesssim10^2$, and $\gamma_0\sim10^2$, we can see that the classical radiation reaction parameter
    \begin{equation}
        R_c=a_0^2\gamma_0 K \sim 10^{-2} \,.
    \end{equation}
    This implies that radiation reaction effects are {relatively} weak, and in particular the radiative energy loss {per laser cycle} should be only a few percent of the initial beam energy. {Nonetheless, radiative losses could be detected with sufficiently long pulses in this regime \cite{Koga:PhysPlas2005,Thomas:PRX2012,Heinzl:2013}.}

    {We find a sizeable angular deflection of the electrons also for small $R_c$.} We may assume that in Eq.~\eqref{eq:angle}, $\mathcal H_1 \sim O(1)$ and $\mathcal I \sim O(L)$, where $L$ is the number of laser cycles. Thus, the largest term in the numerator of Eq.~\eqref{eq:angle} is $\frac{2}{3}R_c \mathcal H_1 a_0 u^{x}_0$, and in the denominator $u^x_{0}\gg \frac{2 R_{c}}{3} a_0\mathcal H_1$. Thus, the expression for the deflection angle $\theta$, Eq.~\eqref{eq:angle}, can be simplified to 
    \begin{align} \label{eq:theta}
        \theta \simeq  \frac{2}{3} a_0^3 K \mathcal H_1 \simeq 
        \frac{2}{3} \frac{a_0^3}{a_\mathrm{crit}}  \mathcal H_1
    \end{align}
    Here we have obtained a simple expression for the electron deflection angle, which essentially depends on $a_0^3$, but surprisingly does not depend on the initial beam energy $\gamma_0$. The angle is also proportional to the pulse asymmetry function $\mathcal H_1$, Eq.~\eqref{eq:Hphi}, which depends only on the laser pulse profile $f_1$, and can have both positive and negative values. Thus, by tailoring the laser pulse we can steer the electron beam to either go upstream or downstream of the laser axis. At first glance, the possibility to achieve positive and negative deflection angles seems counterintuitive, since in the latter case the electrons after the interaction propagate opposite to the laser propagation direction.

    With the help of Eq.~\eqref{eq:theta} we can estimate the size of the deflection angle. Let's say we want to achieve a deflection angle of $1$ mrad, which should be possible to detect experimentally, and assuming again that the asymmetry function is $\mathcal H_1 \sim O(1)$, we find
    \begin{align}
        a_0 \sim \frac{\sqrt[3]{a_\mathrm{crit}}}{10} \approx 36\,.
    \end{align}
    We emphasize again that this effect is purely due to radiation reaction effects and completely ceases to exist if radiation reaction is not taken into account. Moreover, as we have shown sizeable angles on the order of milliradians are achievable even if radiation reaction parameter $R_c=a_0^2\gamma_0K\approx 2.8\times 10^{-5}\gamma_0$ is not large. Since value of the deflection angle is independent of $\gamma_0$ we can in principle keep $R_c$ very small while maintaining large $\theta$. Keeping $\gamma_0$ small enough is also relevant to stay within the realm of applicability of the Landau-Lifshitz equation as a classical description of radiation reaction effects. The latter ceases to be valid if the quantum parameter $\chi = a_0\gamma_0\omega_0/m =  a_0 \gamma_0 K/\alpha \approx 3\times10^{-6} a_0\gamma_0 $ approaches unity.

    \section{Analysis of different pulse shapes}
    \label{sect:pulseshapes}

    Having established the principal signature of radiation reaction, we will now analyze analytically different classes of laser pulse profiles $f$ that can provide a large asymmetry to achieve sizeable values of $\mathcal H_1$. To ensure the property of non-unipolarity of the pulses we define directly the pulse shapes of the vector potential and obtain the electric field shapes by differentiation. We will stay within the classical theory and the plane wave model for the laser pulse. The validity of these assumptions will be investigated below in Section~\ref{sect:sim} with help of numerical simulations.

    \subsection{Few cycle pulses with CEP control}

    The first class of laser pulses profiles that lend themselves to a deflection angle control are few-cycle pulses with controllable carrier-envelope-phase (CEP) $\phi_{CE}$. Here, we parameterize them as
    \begin{align} \label{eq:f-single}
        f(\phi) = g(\phi) \cos (\phi+\phi_{CE} ) \,,
    \end{align}
    with an envelope function
    \begin{align} \label{eq:g}
        g(\phi) & = \begin{cases}
            \cos^4 \left( \frac{\phi}{2L} \right)\,, & -\pi L<\phi<\pi L \,,\\
            0\,, & \text{elsewhere}\,.
        \end{cases}
    \end{align}
    The benefit of this functional form is that all the integrals $\mathcal I$ and $\mathcal H_1$ can be performed analytically. The reason for using a $\cos^4$ instead of the more often used $\cos^2$ envelope has its origin in the numerical simulations we are performing, and details will be given below in the appropriate section. Since the resulting expressions are lengthy they are not shown here in their entirety. For the pulse envelope given in Eq.~\eqref{eq:g}, the relation between the parameter $L$ and the FWHM of the field is given by $T_\mathrm{FWHM}[\unit{\fs}] = 1.21 \, \lambda_0[\unit{\um}]\, L$, where $\lambda_0$ is the laser wavelength.

\begin{figure}[!t]
        \centering
            \includegraphics[width=\columnwidth]{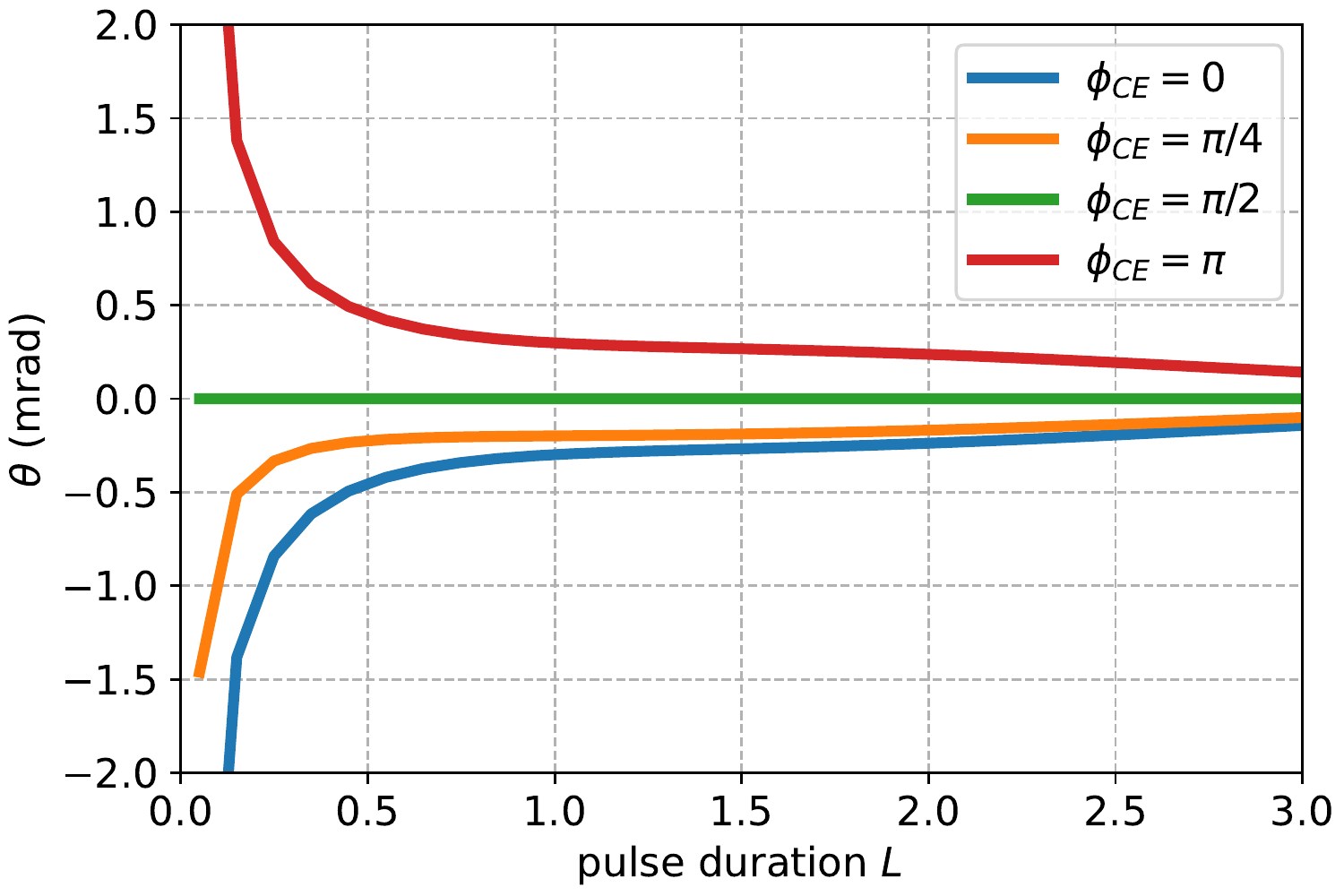}
         \caption{Electron deflection angle $\theta$ as a function of pulse duration $L$ for various CEP values, for $a_0=30$.}
         \label{fig:cep}
    \end{figure}

    The pulses described by Eq.~\eqref{eq:f-single} are quite simple pulse shapes which allow only a limited amount of control of the asymmetry function $\mathcal H_1$. 
    For instance, we find as an analytical result for $L=1$,
    \begin{align}
        \mathcal H_1 = -\frac{237 \pi}{1024} \cos \phi_{CE} - \frac{15 \pi}{2048} \cos 3 \phi_{CE} \,.
    \end{align}
   By changing $\phi_{CE}$ we can achieve both positive or negative values of $\mathcal H_1$, and hence positive or negative deflection angles scattering the electrons upstream or downstream of the laser pulse. For negative values of $\theta$ the $z$-component of the electrons' velocity is negative which means they are deflected upstream of the laser. The electron deflection angles for various CEP are shown in Fig.~\ref{fig:cep} as function of the pulse duration $L$. The analytical calculations in the plot were performed for parameters  $a_0=30$, $\omega_0=1.55\, \unit{\eV}$ and $\gamma_0=195.7$ (i.e.~100 MeV electron energy). It is clear that sizeable deflection angles on the mrad level can only be achieved for ultra-short (single- or sub-cycle pulses) with $L\lesssim 1$. For $\phi_{CE}= n\pi$, with integer $n$ the magnitude of the scattering angle is maximized, and for $\phi_{CE}=(n+1/2)\pi$ the deflection is identically zero. With increasing $L$ the deflection angles quickly decrease to very small values for pulses with more than 3 laser cycles.

    \subsection{Two-color pulses}

    As a second class of laser pulses that provide a pulse asymmetry $\mathcal H_1$ we discuss two-color laser pulses.
    {Two-color pulses are proposed to be useful in a large number of important laser-matter interaction applications, such as: laser plasma acceleration \cite{yu_two-color_2014,PhysRevLett.114.084801,pathak_all_2018,schroeder_two-color_2018,li_laser-plasma_2019}, high-harmonic generation \cite{Yeung2016}, spin-polarized lepton beams \cite{Chen:PRL2019,seipt_ultrafast_2019}, etc.
    We define the two-color wave form as
    }
    \begin{align} \label{eq:2color}
        f(\phi) = \frac{g(\phi)}{N} \Big\{ \cos \phi +c_2 \cos \big[ 2 (\phi+\phi_{2} )\big] \Big\} \,
    \end{align}
    where $c_2$ is the relative amplitude of the second color, and $\phi_2$ denotes the relative phase between the two colors. The normalization factor $N$ is determined in such a way that all pulses have the same value $\mathcal I(\phi_f)=const.$ for all parameters $\phi_2$, $c_2$.

\begin{figure}[!t]
        \centering
        \includegraphics[width=\columnwidth]{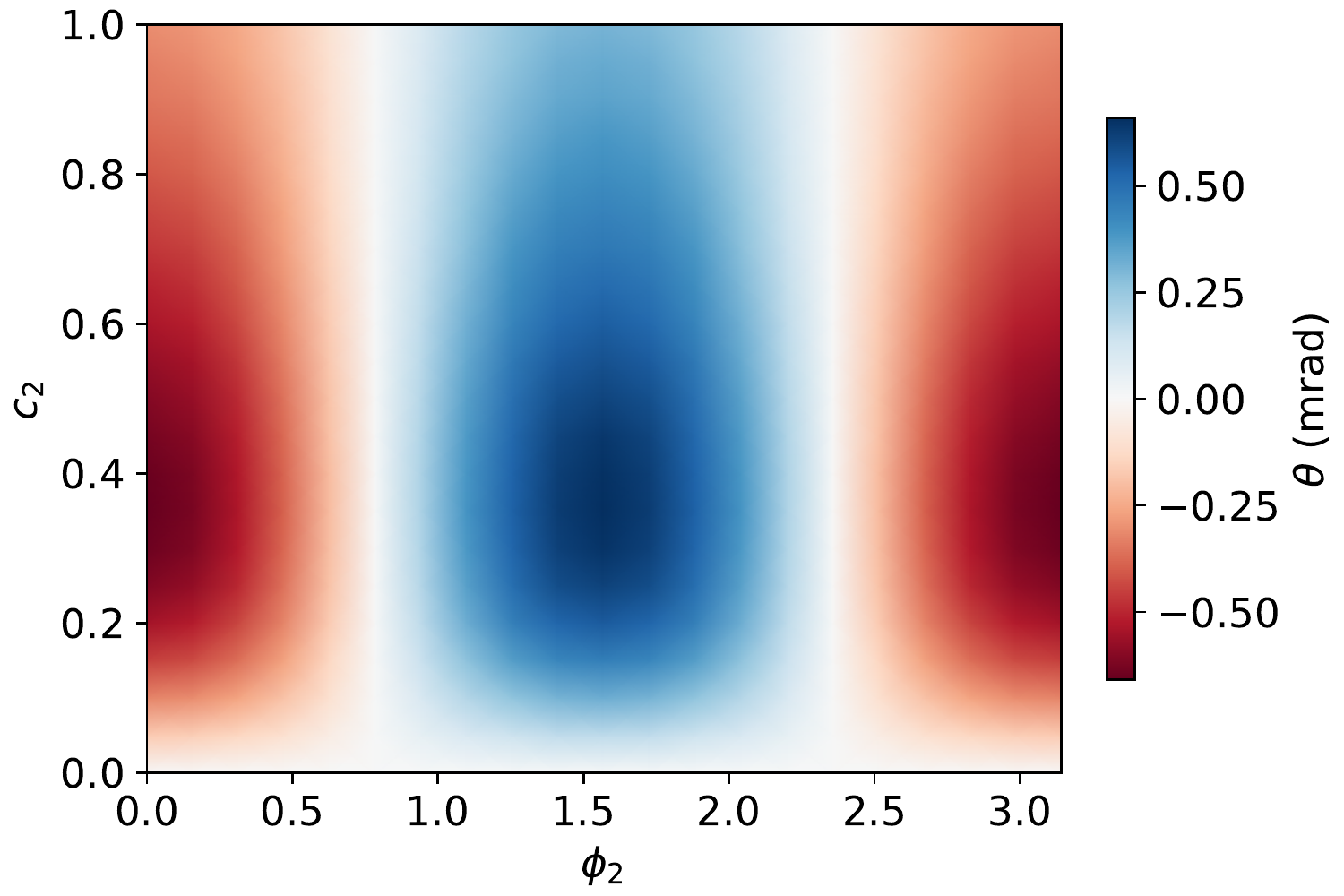}
        \caption{Angular deflection $\theta$ in a two-color pulse for $a_0=30$ and $L=8$.}
        \label{fig:two-color}
    \end{figure}

    By adding a second color with controllable relative phase we can effectively increase the value of $\mathcal H_1$ and achieve additional control over its sign and magnitude. The results for the deflection angle in a two-color pulse are shown in Fig.~\ref{fig:two-color} for $a_0=30$ and $L=8$ as function of $c_2$ and $\phi_2$. For $\phi_2=\pi/4$ the pulse asymmetry vanishes, and this yields a vanishing deflection angle. By changing the relative phase $\phi_{2}$, positive or negative deflection angles can be obtained, which maximize at $\phi_2=\pi/2$ and $0$, respectively. An amplitude of $c_2\simeq 1/3$ maximizes the deflection angle. Calculations for different pulse duration $L$ have shown that the scattering angels also increases by increasing $L$. 

    \subsection{Pulses with a sub-harmonic contribution}

\begin{figure}[!t]
    \centering
    \includegraphics[width=\columnwidth]{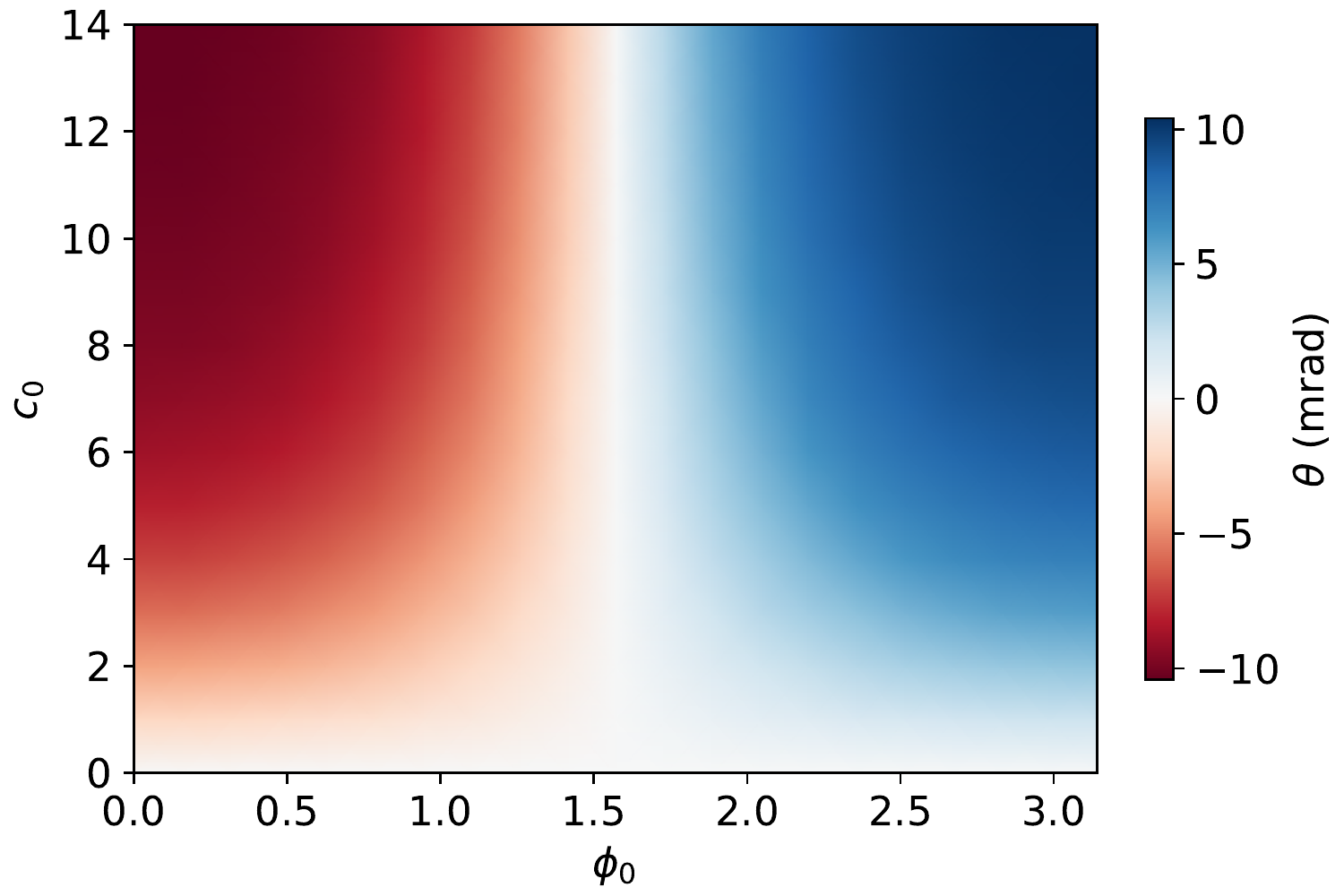}
    \caption{Angular deflection for a pulse with sub-harmonic admixture for $a_0=30$ and $L=8$.}
    \label{fig:subharmonic}
    \end{figure}

    As a third class of pulses we consider the addition of some low-frequency component,
    \begin{equation} \label{eq:subharmonic}
        f(\phi)=\frac{g(\phi)}{N} \left\{ \cos \phi + c_0 \cos  \left( \frac{\phi}{2L}+\phi_{0} \right) \right\},
    \end{equation}
    where $c_0$ is the relative amplitude of a low-frequency sub-harmonic that has a wavelength similar to the pulse envelope duration, $\phi_0$ is the relative phase between the two components. The additional pulse here is essentially a half-cycle far-infrared or THz pulse. The normalization constant $N$ is again chosen to keep $\mathcal I(\phi_f)=const.$ upon varying $\phi_0$ and $c_0$. Adding such a sub-harmonic to the fundamental can give vector potentials which are always positive (or negative) throughout the pulse. {Nonetheless, Eq.~\eqref{eq:subharmonic} describes a non-unipolar pulse as required.} The larger asymmetry {provides larger values of $\mathcal H_1$ and, thus,} to increased deflection angles, as can be seen in Fig.~\ref{fig:subharmonic}, where the typical angles are $\mathcal O(10\, \unit{\milli\radian})$ instead of $\mathcal O(1\, \unit{\milli\radian})$ in the previous cases. The deflection angles are maximized for phases $\phi_0 = 0$ or $\pi$. Moreover, over the studied range the deflection angles monotonically increased with the amplitude $c_0$.

    \section{Numerical Simulations}
    \label{sect:sim}

    So far, we have investigated the novel signatures of radiation reaction in sidescattering based on the classical Landau-Lifshitz equation and a plane wave model for the laser pulse. The Landau-Lifshitz equation describes radiation reaction completely deterministically as a friction force. In reality, the photon emission is quantized, introducing stochasticity in the process. Moreover, in order to achieve the values of $a_0\sim 30$ required for sufficiently large deflection angles, the laser pulses have to be tightly focused.

    In the following we take these effects into account by performing particle-in-cell (PIC) simulations of the electron-beam laser interaction using the code SMILEI \cite{Smilei}. The radiation emission is simulated using the Monte-Carlo radiation model of quantum radiation reaction \cite{blackburn,niel,Neitz:PRL2013,fedotov_advances_2023}. Due to the stochasticity of radiation reaction, here we obtain deflection angle distributions, the mean value of which should correlate with the classically calculated deterministic deflection angle for small values of the quantum parameter $\chi$ \cite{niel,Ridgers:JPP2017}.

    Here it becomes evident why we use the $\cos^4$-shaped pulse envelope in our modeling: The $\cos^2$ pulses are just not smooth enough at the boundary. Since we specify a non-unipolar field where the laser vector potential  $\vec A \propto g$, the electric field contains a term $\propto g'$, which is not smoothly differentiable at the boundary of the pulse for $g\sim\cos^2$. The FDTD method employed in the particle-in-cell code \cite{Smilei} for propagating the electromagnetic fields requires smooth derivatives of $\vec E$ and $\vec B$, though, otherwise the field solver produces numerical artifacts.

    \subsection{1D simulations}

    We have first performed 1D simulations in order to compare our analytical predictions from the Landau-Lifshitz equation with the stochastic Monte Carlo emission model. Such a comparison is best done in a 1D geometry since here the electromagnetic fields are actually plane waves. Any 3D effects due to the focusing of the laser (e.g. ponderomotive forces) and geometric overlap between the laser focus and the electron beam would make a clear comparison much harder. These effects are investigated separately using full scale 3D simulations below. 

\begin{figure}[!th]
    \begin{center}
        \includegraphics[width=\columnwidth]{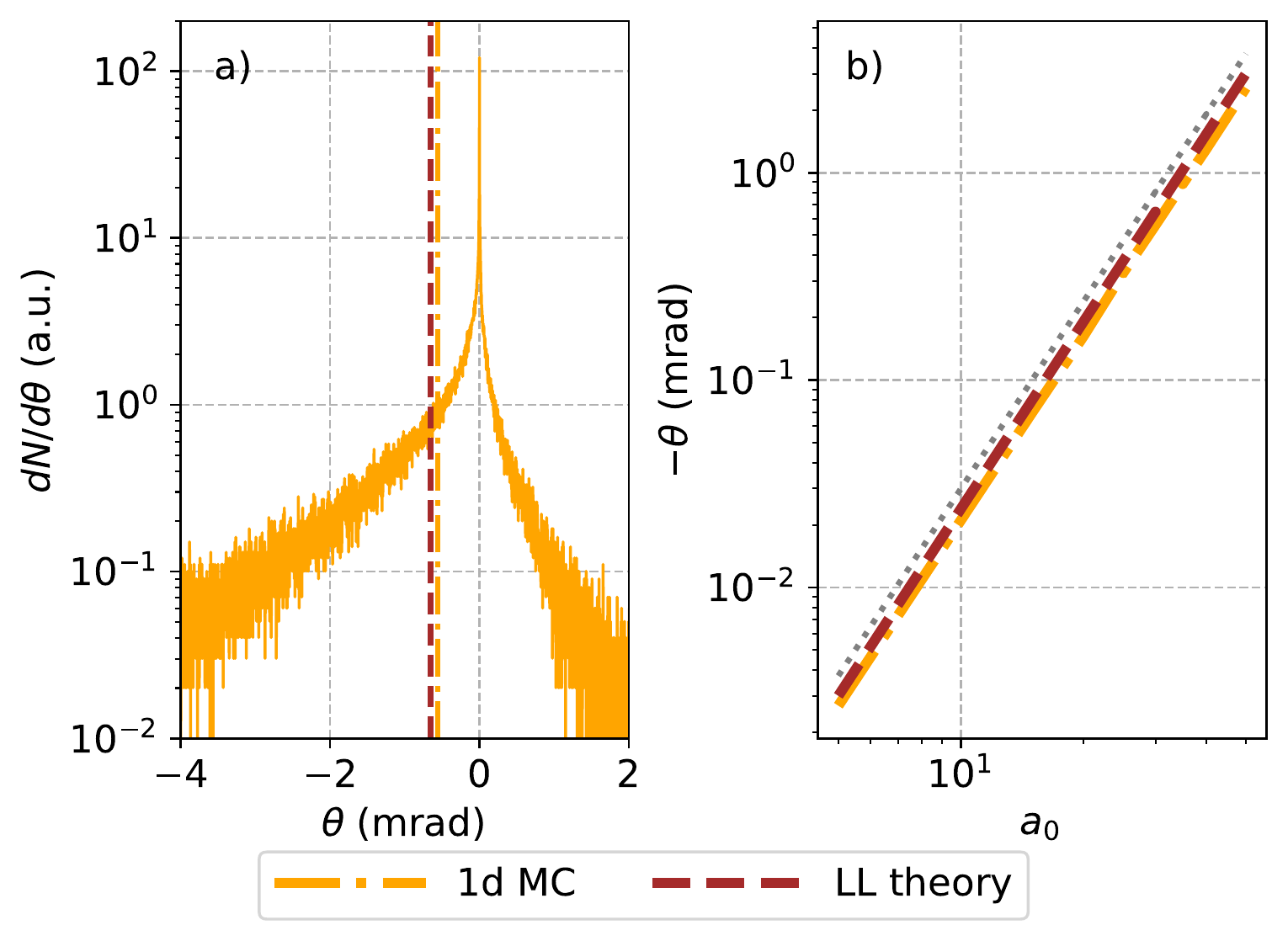}
    \end{center}
    \caption{Deflection angle distribution $dN/d\theta$ from a 1D Monte Carlo simulation (left panel) for $a_0=30$. Right panel: Mean deflection angle and analytical predicted angle as function of $a_0$. $L=8$ for both panels.}
    \label{fig:1d}
    \end{figure}

    The results of our 1D simulations are shown in Fig.~\ref{fig:1d}. For the numerical simulations we have considered only the case of two-color pulses with the analytically found optimal values of $c_2=1/3$ and $\phi_2=0$. The laser pulse is initialized in the code by specifying the transverse magnetic field \cite{Smilei} as the analytically calculated shape function $f'$, where $f$ is given by Eq.~\eqref{eq:2color}. The electrons are initialized as a low-density plasma in the center of the simulation domain with a transverse fluid velocity $v_x=\sqrt{1-1/\gamma_0^2}$ corresponding to a particle energy of $100$ MeV.

    To compare the simulation results with the analytic theory we calculated the mean value $\bar \theta$ of the electron deflection angle distributions $dN/d\theta$. The left panel shows $dN/d\theta$ for a simulation with $a_0^3$ and $L=8$. In addition this plot also contains the mean $\bar \theta$ and the prediction from the Landau-Lifshitz theory as orange and brown dashed vertical lines. The average of the MC is slightly smaller than the classical LL prediction, but the difference between the two is small.
    The right plot of Fig.~\ref{fig:1d} shows a comparison of the classical LL prediction and the MC average over a wide range of $a_0$. Throughout the whole region the difference between the two curves remains small, and both results scale as $\propto a_0^3$. To guide the eye, the gray dashed curve being exactly proportional  to $a_0^3$ was added to the plot.

    {We have also investigated the influence of the electron beam energy spread and emittance on the deflection angle. The analytical calculation predicts that the deflection angle is independent of the electron energy, and this is what we found also numerically. To see this explicitly, we have performed several simulations where the electron energy spread---between zero and up to $\Delta\gamma/\gamma_0 = 30\,\%$---has been taken into account. We found that the mean deflection angle changes by less than 1 \% without any clear trend. We believe these variations are due to the statistical fluctuations from the Monte Carlo emission model.} This is in agreement with the classical Landau-Lifshitz model and shows that effects due to a variation of $\chi$ due to the energy spread play only a minor role for our parameters.

    In addition, we studied the influence of the electron beam divergence by giving the electrons in the initial distribution some small incidence angles, $\Theta_{y,z} = p_{y,z}/p_x$, with various rms values $\bar \Theta \lesssim 2 \, \unit{\milli\radian}$. This was implemented by giving the particle distribution a normalized transverse temperature of $T_{y,z}\approx\bar \Theta^2 \gamma_0^2$. In the studied range, the mean deflection angle was independent of $\bar \Theta$. However, the beam divergence did increase the width of the scattered electron angular distribution, i.e.~the variance of $dN/d\theta$.

    \subsection{3D simulations}

    To establish our novel RR signatures for more realistic scenarios we have also performed full scale 3D simulations of the electron-beam laser interactions \cite{Smilei}. {The simulation box has size $\ell_x=56\,\unit{\micro\meter}$, $\ell_y=28.8\,\unit{\micro\meter}$, $\ell_z=28.8\,\unit{\micro\meter}$ with Silver-Muller open boundary conditions \cite{Smilei}. The resolution was $\Delta x=400\,\unit{\nano\meter}$, $\Delta y=100\,\unit{\nano\meter}$, $\Delta z=25\, \unit{\nano\meter}$ with 16 particles per cell, and $\Delta t=80\,\unit{\atto\second}$.} These simulations allow us to investigate the influence of the laser focusing, including spatial variations of $a_0$ and ponderomotive scattering, as well as geometric overlap effects. For the latter we are mainly concerned about the size of the electron beam in relation to the laser focal spot size and pulse duration. For a realistic collision not all electrons will interact with the highest focused intensity and this reduces the achievable deflection angles. The theoretical model presented in Section~\ref{sect:theory} can be seen as an idealized case where all electrons interact with the peak laser intensity.

\begin{figure}[!th]
        \centering
        \includegraphics[width=\columnwidth]{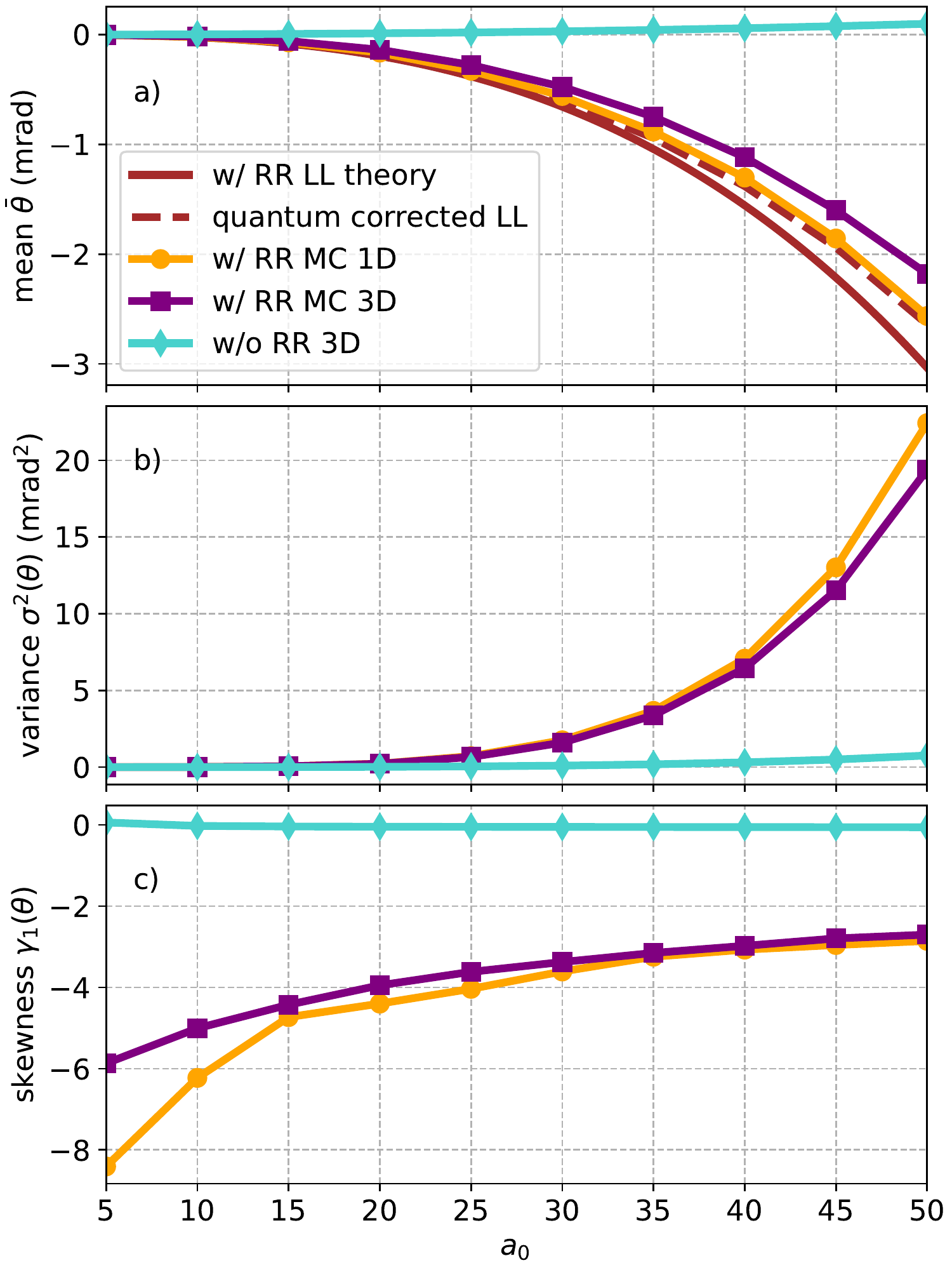}
        \caption{Comparison of the the mean value (a), variance (b) and skewness (c) of the electron deflection angle distributions as function of $a_0$ for different simulation set-ups: 3D simulations with stochastic (Monte Carlo, MC) radiation reactions effects (purple curves), 1D simulations with MC radiation reaction effects (orange curves),
        3D simulations without radiation reaction effects (cyan curves). For the mean angle we also plot the analytical Landau-Lifshitz theory, Eq.~\eqref{eq:angle} (brown curves).}
        \label{fig:a0Scan}
    \end{figure}

    In Fig.~\ref{fig:a0Scan} we present a comparison of 1D and 3D simulations, as well as the case where radiation reaction effects are turned off, i.e. any angular deflection is due to ponderomotive scattering effects. We compare the first three moments of the deflection angle distribution: the mean angle $\bar \theta$, its variance and the skewness.

    In the 3D simulations, the electron beam has a finite transverse rms spot size $\sigma_{y,z}=1\,\unit{\um}$ and finite temporal duration $L_\mathrm{ebeam}=2\,\unit{\fs}$, both modeled as gaussian. The laser temporal pulse profile is again chosen as a two-color pulse with $L=8$, $c_2=1/3$ and $\phi_2=0$. The electron beam is timed to arrive at the laser focus at peak intensity. The transverse beam profile is gaussian with $w_0=10\,\unit{\um}$. 

    As we can see from the top panel of Fig.~\ref{fig:a0Scan}, the behavior of the mean deflection angle is similar for all curves with radiation reaction effects taken into account. As expected, the angular deflection for the 3D simulations is smaller than the 1D results and the analytical Landau-Lifshitz theory prediction. In all cases, the mean deflection angle increases with $a_0^3$, as predicted by the classical LL theory, Eq.~\eqref{eq:theta}. For $a_0 > 35 \ldots 40$, the scattering angle is $\bar \theta >1 \, \unit{\milli\radian}$ for all cases. {Here we also compare the simulation results with the quantum corrected Landau-Lifshitz equation that phenomenologically takes into account the reduction of radiated power due to quantum effects by replacing $\tau_R\to \tau_R g(\chi)$ in Eq.~\eqref{eq:LL}, where the Gaunt factor $g$ is the ratio of the quantum and classical radiation powers \cite{niel,Ridgers:JPP2017}. The deflection angle calculated from the quantum corrected LL equation nearly agrees with the 1D simulations. In agreement with the literature, e.g.~Refs.~\cite{niel,Ridgers:JPP2017}, the main quantum effect at $\chi<0.03$ is the reduction of radiated power, and not so much the stochasticity of photon emission.}

    Part of the reduction in the 3D case can be attributed to the ponderomotive scattering at the laser intensity gradient in the focal spot. The cyan curve represents our 3D simulations in which the radiation reaction model was turned off and hence the angular deflection is solely due to ponderomotive scattering effects (recall that no deflection can occur in the plane wave model). The angular deflection due to ponderomotive scattering has opposite sign than the radiation reaction effect. Moreover, the case without radiation reaction scales as $\propto a_0^2$ as one would expect for an effect of the ponderomotive force $\vec F_\mathrm{pond}\propto -\vec \nabla I_\mathrm{laser}$.

    {Panel b)} of Fig.~\ref{fig:a0Scan} shows that the deflection angle distributions become much broader due to the effect of radiation reaction in comparison to the no-RR case. The variance strongly increases with $a_0$. The difference between the 1D and 3D simulations is much smaller than the difference between the cases with and without RR. {A plot of the skewness in Fig.~\ref{fig:a0Scan} c)} shows that the scattering angle distribution is approximately symmetric without radiation reaction, but becomes quite asymmetric when radiation reaction is taken into account.

    \subsection{Systematic Study of Collision Parameters within the 3D model}

    We now investigate the dependence of the deflection angle on various important laser and electron beam parameters. For all simulations we use $L=8$ (except when stated otherwise), $c_2=1/3$ and $\phi_2=0$.

\begin{figure}[!bth]
    \centering
    \includegraphics[width=\columnwidth]{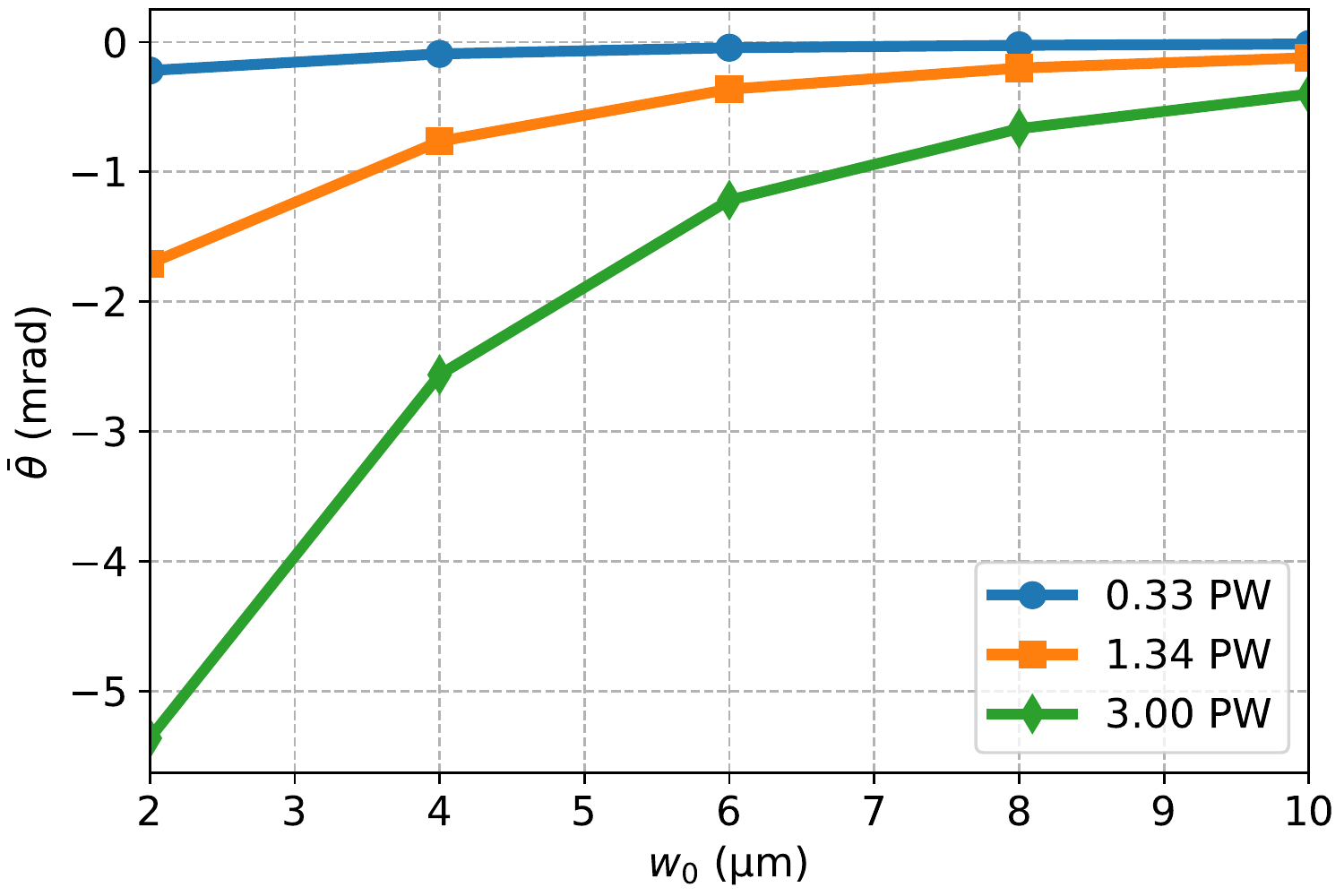}
    \caption{Mean electron deflection angle as function of laser focal spot size $w_0$ for constant laser power. 
    Electron beam parameters are ${\sigma_y}=\sigma_z = 1 \, \unit{\um}$ and $L_\mathrm{ebeam}=12\, \unit{\fs}$.}
    \label{fig:ConstPower}
    \end{figure}

    In Figure~\ref{fig:ConstPower} we keep the laser power constant and determine the mean scattering angle as a function of $w_0$. 
    As the laser is focused more tightly, the value of $a_0$ increases and hence the electron deflection angle. If we assume 1~mrad deflection as the threshold for detection, we see that the case with 330 TW is not sufficient to see the radiation reaction signal.
    With a (multi-)PW laser, the deflection angles can be significantly larger than 1~mrad even for larger spot sizes $w_0$.

\begin{figure}[!h]
      \centering
        \includegraphics[width=\columnwidth]{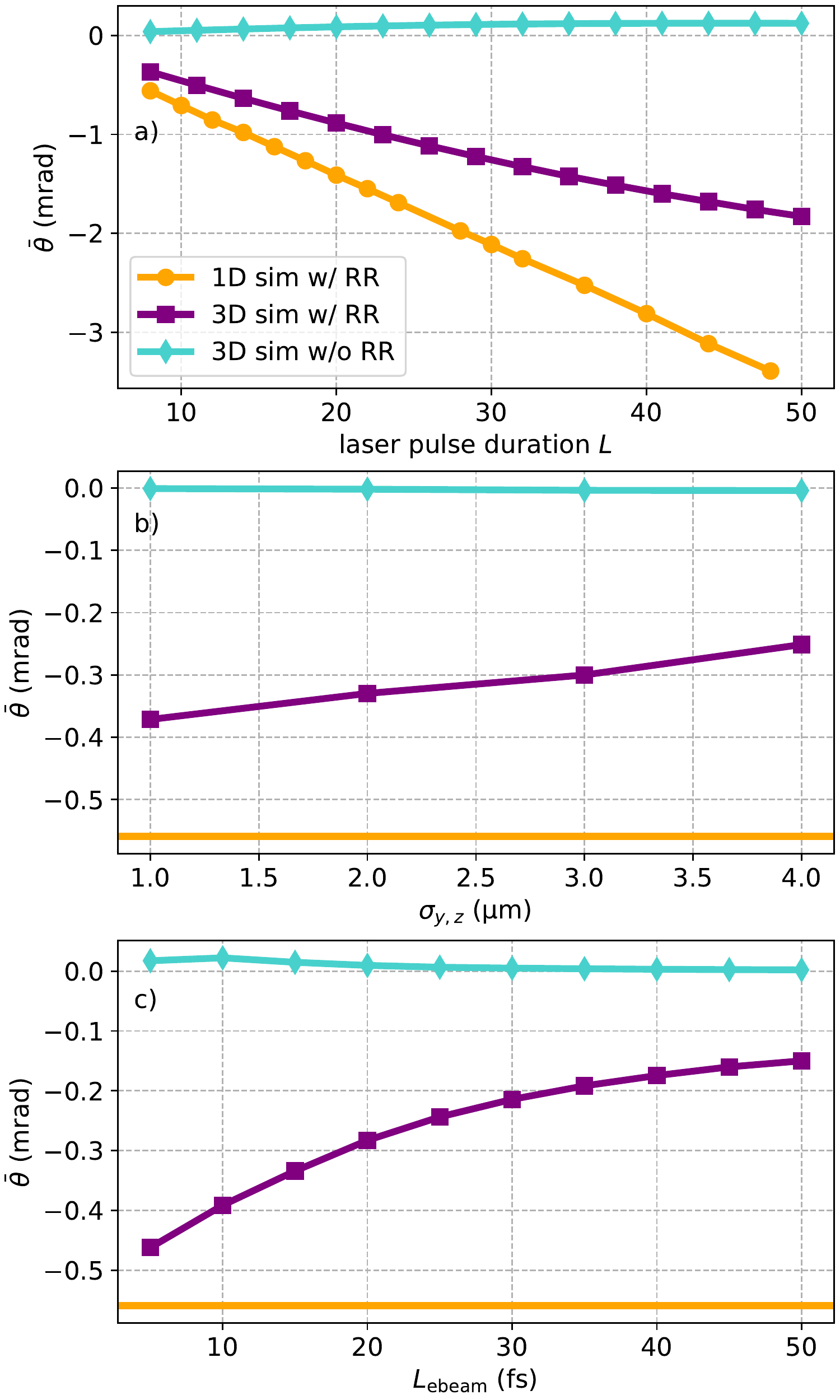}
     \caption{{Dependence of the electron deflection angle on different beam parameters influencing the geometric overlap of the laser and electron beam. Depicted are the dependencies on the laser pulse duration (a), electron beam transverse size (b) and the electron beam duration (c).}}
     \label{fig:overlap}    
     \end{figure}

    The Figure~\ref{fig:overlap} we plot results showing the influence of the geometric overlap of the laser and the electron beam. The focal spot size $w_0=10\, \unit{\um}$ and $a_0=30$ are kept constant for all simulations.

    The top panel of Fig.~\ref{fig:overlap} shows the dependence on the laser pulse duration $L$, while the electron beam size is kept constant at ${\sigma_y}=\sigma_z=1\, \unit{\um}$ and $L_{\mathrm{ebeam}} = 12\, \unit{\fs}$. For the 1D simulations with radiation reaction, the scattering angle increases linearly with $L$ following the prediction from the analytical LL theory. For the 3D simulation the angle grows slower and seems to saturate for large $L$. This behavior is due to a limitation of the effective interaction time, which for long laser pulses becomes effectively equal the focal spot transversal time proportional to $w_0$. Thus, extending the pulse duration beyond that level is not efficient. For comparison we also show results of 3D simulations without radiation reaction effects which show very small deflection angles (and with opposite sign) for all $L$. {For comparison, we also performed one 3D simulation without the second color in the laser pulse which yielded a much smaller deflection angle of $\bar \theta \approx -0.14 \,\unit{mrad}$ at $L=30$.}

    In the middle panel of Fig.~\ref{fig:overlap} we show the dependence on the transverse electron beam size ${\sigma_{y,z}}$. Here we keep the beam duration constant at $L_{\mathrm{ebeam}} = 12\,\unit{\fs}$ and $L=8$. By increasing ${\sigma_{y,z}}$ between 1 and 4 $\unit{\um}$ the deflection angle slightly decreases from $0.37$ to $0.25$ mrad (purple curve).

    The bottom panel of Fig.~\ref{fig:overlap} shows the dependence on the electron beam duration $L_\mathrm{ebeam}$, while keeping constant ${\sigma_y}=\sigma_z=1 \, \unit{\um}$ and $L=8$. Here, the dependence is much more severe than the influence of the transverse spot size. The deflection angle decreases significantly for longer beam duration, which is mainly due to $L_\mathrm{ebeam}$ becoming larger than the effective laser pulse duration. In that case only a fraction of the electrons interact with the laser pulse which decreases the mean scattering angle when averaging over the final beam angular distribution.

\begin{figure}[!th]
    \centering
    \includegraphics[width=\columnwidth]{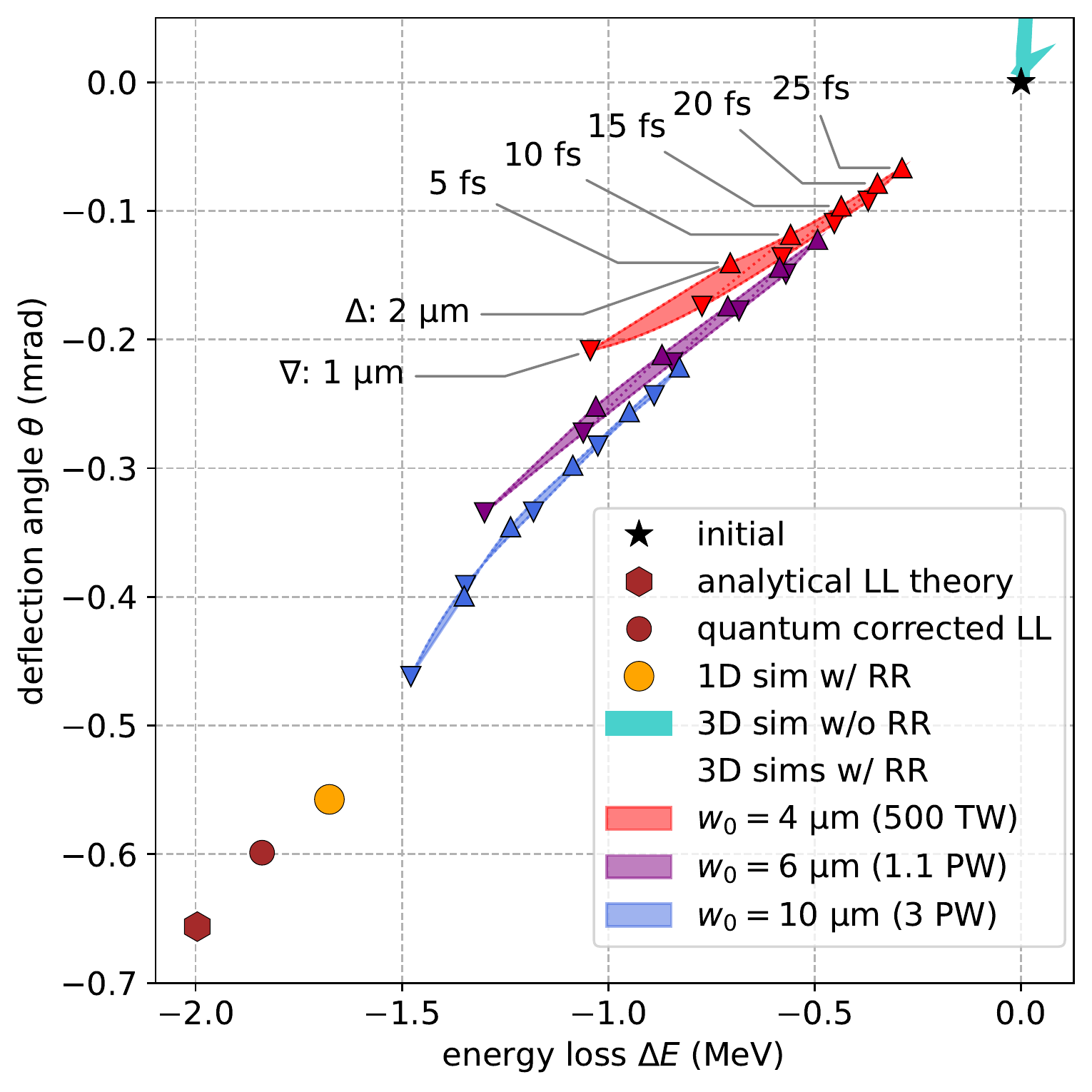}
    \caption{Survey of expected signals of radiation reaction in 90 degree sidescattering over simulations with various parameters.
    The brown hexagon and orange circle symbols are for the analytical LL theory and 1D simulations, respectively.
    The shaded areas represent two-dimensional parameter scans using 3D simulations across transverse electron beam size ${\sigma_{y,z}}$ and beam duration $L_\mathrm{ebeam}$. All results are for $a_0=30$. For the 3D simulations the focal spot size was varied, which equates to variable laser powers of $500\,\unit{\tera \watt}$, $1.1\,\unit{\peta \watt}$ and $3\,\unit{\peta \watt}$, respectively.
    }
    \label{fig:GammaTheta}
    \end{figure}

    With these detailed studies on the various collision parameters done, we now present in Figure~\ref{fig:GammaTheta} a comprehensive survey of the expected signals of radiation reaction in 90 degree sidescattering: the electron beam energy loss $\Delta E$ and the electron deflection angle $\theta$. Before going into the details, we can emphasize that the outcome of the various radiation reaction models significantly deviate from the \nil{}-results (black star, cyan area) in the top right corner of the plot.
    The initial beam is represented by the black star at coordinates $(0,0)$, which is also the outcome for a 1D plane-wave model without radiation reaction. Right next to this, in the upper right corner the cyan area represents the outcome of 3D simulations without radiation reaction.

   The result of the analytical Landau-Lifshitz theory is given by the brown hexagon in the lower left corner of the plot; the 1D simulations are given by the orange circle with slightly smaller deflection angle and energy loss.  The red, purple and blue areas represent two-dimensional parameter scans using 3D simulations for transverse electron beam size ${\sigma_{y,z}}=1\dots2\,\unit{\um}$ and beam duration $L_\mathrm{ebeam}=5\ldots25\,\unit{\fs}$. Downward pointing triangles are for ${\sigma_{y,z}}=1\,\unit{\um}$, while upward pointing triangles represent simulations with ${\sigma_{y,z}}=2\,\unit{\um}$.  For the survey we keep constant $a_0=30$, and the initial electron beam energy of $104$ MeV. All models and simulations with radiation reaction show negative angles and a non-negligible energy loss for all the parameters, while for the case without radiation reaction effects the angle is positive and no energy loss. 

    From the plot we can clearly observe the trends for the various parameters: It is favorable to have a larger laser spots $w_0$, i.e. more laser power (blue is better than red) and shorter electron beams (5 fs symbols are farther away from the \nil{}-result than the 25 fs symbols in each category). Moreover, the electron beam transverse size has a much smaller effect on the deflection angle and energy loss.

    \subsection{An Indirect Signal of Laser Depletion}

\begin{figure}[!tbh]
        \centering
        \includegraphics[width=\columnwidth]{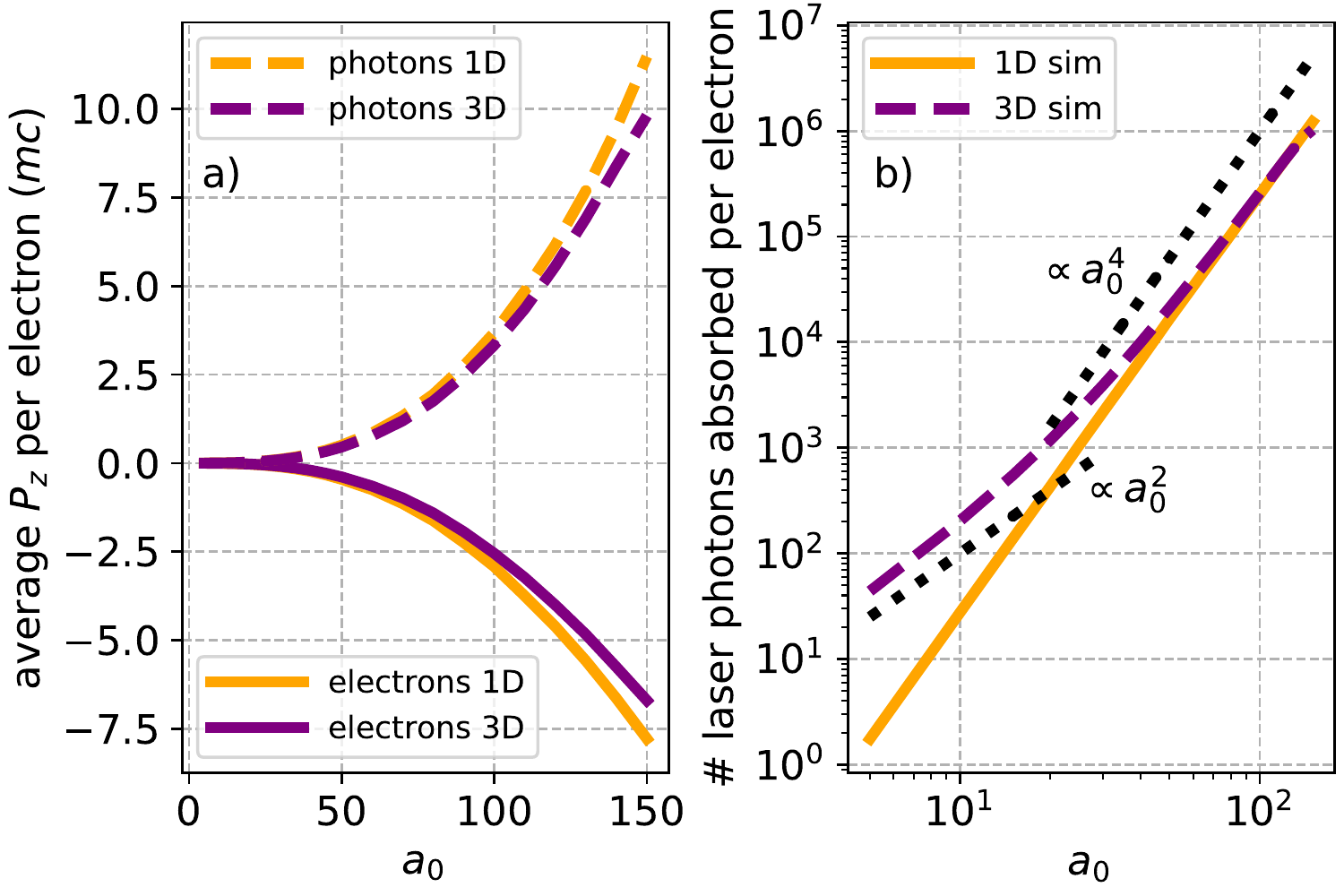}
        \caption{In a 90 degree scattering geometry, measuring the momentum component along the laser beam axis for all final state particles can provide an indirect measure for laser photon absorption.}
        \label{fig:depletion}
    \end{figure}

    When measuring the total longitudinal momentum of the scattered electrons \emph{and} emitted photons we can make some (indirect) statements about laser depletion during the strong-field QED interaction \cite{Seipt:PRL2017,Ilderton2018,heinzl_mode_2018} exploiting total momentum conservation, $P_i = P_f$.
    Microscopically, quantum radiation reaction is described as multiple nonlinear Compton scattering events, in each of which multiple laser photons are absorbed from the electron while a single high-energy photon is emitted.
    When comparing the {$P_z$} momentum components of the electrons and {high-energy} photons after the scattering, normalized to the number of initial electrons, we observe an imbalance in both 1D and 3D simulations, see Fig.~\ref{fig:depletion}.
    The sum of the {$z$}-components of final electron momentum and the momenta of all the photons it has emitted during the interaction are non-zero. Since in the initial state the electrons are propagating at 90 degrees with respect to the laser axis their initial momentum {$P_{e,z}(t\to-\infty)=0$}. This means that the final {$z$}-momentum must come from the absorption of laser photons.

    {The number of laser photons absorbed per incident electron, ${n_A = (P_{\gamma,z} + P_{e,z}) /\omega_0}$ is shown in the right panel of Fig.~\ref{fig:depletion}. In 1D simulations this quantity scales proportional to $a_0^4$, as expected from theoretical considerations for laser depletion during radiation emission~\cite{Ritus:JSLR1985,Seipt:PRL2017}. For $a_0\gtrsim 100$ the average number of absorbed laser photons per incident electron reaches the order of $10^6$. The 3D simulations show the same behavior for large $a_0$. However, for small $a_0$ ($a_0<30$), when the number of absorbed laser photons becomes small, we find an $a_0^2$-scaling which indicates a dominance of ponderomotive scattering effects over radiation reaction for this observable.}

    \section{Summary and Conclusion}
    \label{sect:summary}

    In this work, we have investigated the radiation reaction effects in electron-beam laser collisions and put forward a novel signature of RR effects; an electron deflection angle along the laser beam axis in the 90 degree side-scattering geometry.
    
    We have derived from the classical Landau-Lifshitz theory the scaling of the scattering angle with $a_0^3$ and it being independent of the electron beam energy. The crucial factor for the angular deflection is a temporal asymmetry in the laser vector potential, expressed via the parameter $\mathcal H_1$ in Eq.~\eqref{eq:Hphi}. By controlling the laser pulse shape the electron beam can be steered to be deflected upstream or downstream of the laser. We analyzed three different types of pulse shapes that can provide large values for the parameter $\mathcal H_1$ and the required controls, with the conclusion that two-color pulses with adjustable relative phase between the two colors seem to be the best candidates in terms of large enough angles and experimental achievability.
    
    We then have performed full scale Monte Carlo simulations of the collision using the PIC code SMILEI in order to assess the effects of stochasticity of the photon emission, laser focusing and related ponderomotive scattering, as well as geometric overlap between the beams. Our results show that these effects can reduce the analytically predicted deflection angles, the values of the RR signals remain clearly distinct from the \nil{}-result, see Fig.~\ref{fig:GammaTheta}. Thus, the electron deflection studied in this paper could serve as a novel signature of radiation reaction effects with present-day (multi-)PW class lasers and high-quality LWFA electron beams. Further, a simultaneous measurement of the $\gamma$-ray yield could provide an indirect measurement of laser depletion during the scattering process for $a_0 \gtrsim 30$.

    Some effects such as shot-to-shot timing and beam pointing jitter have not been addressed in this paper. However, they need to be well under control experimentally for a successful verification of the RR signatures predicted in this paper \cite{maier_decoding_2020,Huang_2022,seidel_polarization_2024}. We expect that electron beam pointing jitter would need to be $\Delta\theta_b \ll 0.1$~mrad, and that the laser-electron beam arrival-timing jitter would need to be small compared to the pulse duration, in addition to the laser focal distribution not varying significantly relative to the beam. Analysis of the sensitivity to these effects is left for future work.

    \section*{Acknowledgements}

    This work was supported by the US National Science Foundation GACR collaborative grant 2206059 and NSF grant 2108075. 
    SSB was supported by U.S. Department of Energy Office of Science Offices of High Energy Physics under Contract No. DE-AC02-05CH11231.
    The authors gratefully acknowledge the Gauss Centre for Supercomputing e.V. (www.gauss-centre.eu) for funding this project by providing computing time through the John von Neumann Institute for Computing (NIC) on the GCS Supercomputer JUWELS at Jülich Supercomputing Centre (JSC). The research leading to the presented results received additional funding from the European Regional Development Fund and the State of Thuringia (Contract No. 2019 FGI 0013).

	\input{references}

\end{document}